\documentclass[showpacs,a4paper,11pt]{article}
\usepackage{jheppub} 
\usepackage[T1]{fontenc} 

\title{\boldmath ${\cal O}(\alpha_s)$ corrections to the B-hadron energy distribution of the
 top decay in the general two Higgs doublet model  considering GM-VFN scheme}

\author[a,b]{S. Mohammad Moosavi Nejad}

\affiliation[a]{Faculty of Physics, Yazd University, P.O. Box
89195-741, Yazd, Iran}
\affiliation[b]{School of Particles and Accelerators,
Institute for Research in Fundamental Sciences (IPM), P.O.Box
19395-5531, Tehran, Iran}

\emailAdd{mmoosavi@yazduni.ac.ir}

\abstract{We present our analytic results for the NLO corrections to the partial decay width $t\rightarrow H^+b$
followed by $b\rightarrow BX$ for nonzero b-quark mass ($m_b\neq 0$) in the Fixed-Flavor-Number
scheme (FFNs). To  make our  predictions for the energy  distribution of the outgoing bottom-flavored
hadron (B-hadron) as a function of the normalized B-energy fraction $x_B$, we apply the General-Mass
Variable-Flavor-Number scheme (GM-VFNs)  in the general two-Higgs-doublet model.  In order to
describe both the b-quark and the gluon hadronizations in top decay we use fragmentation functions
extracted from data from $e^+e^-$ machines. We find that  the most reliable prediction for the B-hadron energy spectrum
is made in the GM-VFN  scheme.
}

\begin{document} 
\maketitle

\section{Introduction}
\label{sec:intro}

The top quark has been the latest standard model particle discovered by the
CDF and D0 experiments at  Fermilab Tevatron \cite{Group:2009ad}. Its properties
are  essential for our understanding of the standard model (SM) theory. At the Large Hadron Collider (LHC),
one expects a cross section $\sigma(pp\rightarrow t\bar{t}X)\approx 1$   (nb) at design energy
$\sqrt{S}=14$ TeV  \cite{Langenfeld:2010zz}. With the LHC design luminosity
of $10^{34} (cm)^{-2}(sec)^{-1}$, it is expected to produce a $t\bar t$ pair per second.
Thus, the LHC is  a superlative top factory, which  allows  to carry out precision tests of the SM
and, specifically a precise measurement of the top quark properties such as its mass.
Due to the element $|V_{tb}|\approx 1$ of the Cabibbo-Kobayashi-Maskawa (CKM) \cite{Cabibbo:1963yz}
quark mixing matrix, the top quark  is decaying dominantly through the mode $t\rightarrow bW^+$ within
the SM. As it is well known, bottom quarks hadronize before they decay, therefore each b-jet contains a B-hadron
 which most of the times is a B-meson. Events with B-mesons are identified by a displaced decay vertex
associated which charged-lepton tracks. This is precisely the signature used to identify b-jets.
In Ref.~\cite{Kniehl:2012mn}, we studied the  B-meson energy
distribution produced in top decay considering both the bottom quark and the gluon fragmentations.
We also studied  the angular distribution of the W-boson decay products in the decay chain
$t\rightarrow bW^+\rightarrow Bl^+\nu_l+X$. The effects of b-quark and hadron masses are also considered.

In many extensions of the SM such as the minimal supersymmetric standard model (MSSM), the
Higgs sector of the SM is enlarged, typically by adding an extra  doublet of complex
Higgs fields. After spontaneous symmetry breaking,
the two scalar Higgs doublets $H_1$ and $H_2$ yield three physical
neutral Higgs bosons (h, H, A) and a pair of charged-Higgs bosons ($H^\pm$) \cite{Djouadi:2005gj}.
The top quark  is decaying dominantly through $t\rightarrow bH^+$
in a model with two-Higgs-doublet (2HDM)
\cite{Gunion}, providing that the top quark mass $m_t$, bottom quark
mass $m_b$ and the charged-Higgs boson mass $m_H^+$ satisfy $m_t>m_b+m_H^+$.
In this case one expects measurable effects in the top quark decay width and decay distributions
due to the $H^\pm$-propagator contributions, which are
potentially large in the decay chain $t \rightarrow bH^+\rightarrow b(\tau^+\nu_\tau)$.\\
At  LHC, the dominant source of top quarks is $pp\rightarrow t\bar t$ process, therefore  the charged
Higgs boson  has been searched for in the subsequent decay
products of the top pairs $t\bar t\rightarrow H^\pm W^\mp b\bar b$ and $t\bar t\rightarrow H^\pm H^\mp b\bar b$
 when $H^\pm$ decays into $\tau$ lepton and neutrino.\\
In our previous work \cite{MoosaviNejad:2011yp}, we studied the energy spectrum of the inclusive
bottom-flavored mesons in the presence of charged Higgs boson
 by working in the massless scheme or zero-mass
variable-flavor-number (ZM-VFN) scheme where the mass of bottom quark is set to zero from the
beginning. As it was shown,  in the limit of vanishing b-quark mass our results in
two variants of the  2HDM are the same. In the present work, we impose the effect of b-quark mass
on the B-spectrum employing the general-mass variable-flavor-number (GM-VFN) scheme.  As it is shown,
the results will  be different in two variants of the 2HDM and it is found the NLO corrections with $m_b\neq 0$
to be significant.\\
To obtain the total B-hadron energy distribution of the top decay, two contributions due to the decay modes $t\rightarrow bW^+$
 (in the SM) and $t\rightarrow bH^+$ (in the 2HDM) should be summed up.  However,
the contribution of SM is always larger than the one coming from 2HDM \cite{MoosaviNejad:2011yp}, but
there is a clear separation between the decay channels  $t \rightarrow bW^+$ and $t \rightarrow bH^+$  in both
the $t\bar{t}X$ pair production and the  $t/\bar{t}X$ single top production at  the LHC \cite{Ali:2011qf}.
New results of a study on the charged-Higgs bosons in pp collision at a center of mass
energy of $\sqrt{s}=7$ TeV  are reported by the ATLAS Collaboration \cite{ATLAS}.

Finally, since bottom quarks fragment into the B-meson, in order to describe the b-quark non-perturbative fragmentation
some phenomenological hadronization models can be
used. Our treatment at NLO in the GM-VFN scheme is
manifestly based on the factorization theorem of QCD which guarantees that the 
fragmentation functions are universal and subject to DGLAP evolutions \cite{dglap}.
 Relying on the universality of the hadronization mechanism, we can tune such
models to data on B production in $e^-e^+$ annihilation data from CERN LEP1 and SLAC SLC and use
them to predict the B-hadron spectrum in top decay.  The hadronization of the b-quark was considered in the NLO
QCD analysis of top-quark decay in Refs.~\cite{Corcella:1, Corcella:2} and was identified to
be the most important factor of uncertainty
in the determination of the top-quark mass.

This paper is organized as follows.
In Sec.~\ref{sec:one}, we give the parton-level expressions for the NLO QCD corrections to the tree-level
rate of $t\rightarrow bH^+$ in the fixed flavor number (FFN) scheme.
In Sec.~\ref{sec:two}, the scheme of GM-VFN is explained by introducing  the perturbative
 fragmentation function $b\rightarrow b$.
In Sec.~\ref{sec:three}, we present our  hadron-level results working in the  GM-VFN scheme.
In Sec.~\ref{sec:four},  we summarize our conclusions.

\boldmath
\section{Parton level results}
\label{sec:one}
\unboldmath

\boldmath
\subsection{Born level rate of $t\rightarrow bH^+$}
\unboldmath

We consider the decay channel $t\rightarrow bH^+$ in the general 2HDM, where  $H_1$ and $H_2$ are the
doublets whose vacuum expectation values give masses to the down and up type quarks, respectively, and
a linear combination of the charged components of $H_1$ and $H_2$ gives the physical charged Higgs $H^+$.
In a general model with two Higgs doublets  to avoid
tree level flavor-changing neutral currents (FCNC), the generic
Higgs coupling to all quarks has to be restricted.  There are two possibilities for
 the two Higgs doublets to couple to the fermions.
In the first possibility (model I), one of the Higgs doublets ($H_1$) couples to all bosons
and the other one ($H_2$) couples to all the quarks.  In this model, the Yukawa couplings between
 the charged Higgs boson, the top and the bottom quarks are given by \cite{GHK}
\begin{eqnarray}
L_I=\frac{g_W}{2\sqrt{2}m_W}V_{tb}\cot\beta \bigg\{H^+\bar{t}\big[m_t(1-\gamma_5)-
m_b(1+\gamma_5)\big]b\bigg\}+H.c.
\end{eqnarray}
For the vacuum expectation values of $H_1 (\textbf{v}_1)$  and  $H_2  (\textbf{v}_2)$,  we
 have $\textbf{v}_1^2+\textbf{v}_2^2=(\sqrt{2} G_F)^{-1}$ where $G_F$ is the Fermi's constant
  and the ratio of the two values is a free parameter and one can define the angle $\beta$  to
 parameterize it, i.e.  $\tan\beta=\textbf{v}_2/\textbf{v}_1$. The weak coupling factor $g_W$
 is related to the  Fermi coupling constant   by $g_W^2=4\sqrt{2} m_W^2 G_F$.
 In the equation above, $H^+=\cos\beta H_2^+-\sin\beta H_1^+$ is the physical charged Higgs boson. \\
 In the second possibility (model II), the doublet $H_2$ couples to the right-handed up-type
  quarks  ($u_R, c_R, t_R$) and the $H_1$ couples to
 the right-handed down-type quarks. In this model, the interaction Lagrangian would be
\begin{eqnarray}
L_{II}=\frac{g_W}{2\sqrt{2}m_W}V_{tb} \bigg\{H^+\bar{t}\big[m_t\cot\beta(1-\gamma_5)+
m_b\tan\beta(1+\gamma_5)\big] b\bigg\}+H.c .
\end{eqnarray}
The Born amplitude for the process $t\rightarrow bH^+$ can be parameterized
 as \textit{$M_0=\bar{u_b}(a+b\gamma_5)u_t$},
therefore, the tree-level total decay width is given by
\begin{eqnarray}\label{gammatree}
\tilde\Gamma_0=\frac{m_t}{16\pi}\bigg\{(a^2+b^2)\big[1+\frac{m_b^2}{m_t^2}-
\frac{m_{H^+}^2}{m_t^2}\big]+
2(a^2-b^2)\frac{m_b}{m_t}\bigg\}\lambda^{\frac{1}{2}}(1,\frac{m_b^2}{m_t^2},
\frac{m_{H^+}^2}{m_t^2}),
\end{eqnarray}
where $\lambda(x,y,z)=(x-y-z)^2-4y z$ is the  K\"all\'en function and for model $I$, one has
\begin{eqnarray}\label{model1}
a^2+b^2&=&\sqrt{2} |V_{tb}|^2 G_F (m_t^2+m_b^2)\cot^2\beta,\nonumber\\
a^2-b^2&=&-2\sqrt{2} |V_{tb}|^2 G_F (m_b m_t)\cot^2\beta,
 \end{eqnarray}
and for model II
\begin{eqnarray}\label{model2}
a^2+b^2&=&\sqrt{2} |V_{tb}|^2 G_F(m_t^2\cot^2\beta+m_b^2\tan^2\beta),\nonumber\\
a^2-b^2&=&2\sqrt{2} |V_{tb}|^2 G_F (m_b m_t).
 \end{eqnarray}
In the limit of vanishing b-quark mass, the tree level decay width is
discussed in \cite{MoosaviNejad:2011yp}.  Since $m_b\ll m_t$, the b-quark mass can always be safely neglected in model I
but in model II, the left-chiral coupling term proportional to $m_b\tan\beta$ can become comparable
to the right-chiral coupling term $m_t\cot\beta$ when $\tan\beta$  becomes large. Therefore, one can not
naively set $m_b=0$ in all expressions in model II.

In the following,  we explain  the calculation of the NLO QCD corrections
to the Born level decay rate of $t\rightarrow bH^+$ and
we present the parton-level expressions for $d\Gamma(t\rightarrow BH^++X)/dx_B$ at
 NLO in the FFN scheme where $m_b\neq 0$ is considered.

\boldmath
\subsection{Virtual one-loop corrections and counterterms}\label{virtual}
\unboldmath

The virtual corrections  to the $tbH^+$-vertex consists of both infrared (IR) and
ultraviolet (UV) singularities where the ir- and uv-divergences arise from the collinear- and the
soft-gluon singularities, respectively. In our calculation, we adopt
the on-shell mass-renormalization scheme and all singularities are regularized by dimensional
regularization in $D=4-2\epsilon$ space-time dimensions.
 To simplify the  formulas we introduce  the following kinematic variables,
 in the notation of Refs.~\cite{kadeer, Czarnecki:1992ig}, along with some other required  variables
\begin{eqnarray}
p_0&=&\frac{1}{2}(1+R-y),
\nonumber\\
\beta &=&\frac{\sqrt{R}}{p_0},
\nonumber\\
p_3&=&p_0\sqrt{1-\beta^2},
\nonumber\\
p_{\pm}&=&p_0\pm p_3,
\nonumber\\
Y_p&=&\frac{1}{2}\ln\frac{p_+}{p_-},
\nonumber\\
Y_w&=&\frac{1}{2}\ln\frac{1-p_-}{1-p_+},
\nonumber\\
H&=& (a^2+b^2)p_0+(a^2-b^2)\sqrt{R},
\nonumber\\
T&=&p_0(1-x_b)\sqrt{x_b^2-\beta^2},
\nonumber\\
\Phi(x_b)&=&p_0 \big[\sqrt{x_b^2-\beta^2}-\ln\frac{\beta}{x_b-\sqrt{x_b^2-\beta^2}}\big],
\end{eqnarray}
where the scaled masses $R=m_b^2/m_t^2$ and $y=m_{H^+}^2/m_t^2$ are defined. Choosing
 these notations,  the tree-level total width (Eq.~(\ref{gammatree})) is simplified to $\tilde\Gamma_0=m_t H p_3/(4\pi)$.
It is also convenient to introduce the normalized energy
fractions $x_i=E_i/E_b^{max}(i=b,g)$ where $E_b^{max}=m_t p_0$.\\
Considering the above notations, the contribution of virtual corrections into the
differential decay width reads
\begin{eqnarray}
\frac{d\tilde\Gamma^{\textbf{vir}}_b}{dx_b}=\frac{p_3}{8\pi m_t}
\overline{|M^{\textbf{vir}}|^2}\delta(1-x_b),
\end{eqnarray}
where,
$\overline{|M^{\textbf{vir}}|^2}=1/2\sum_{Spin}(M_0^{\dagger} M_{loop}+M_{loop}^{\dagger} M_0)$.
The renormalized amplitude of the virtual corrections is written as
 $M_{loop}=\bar{u_b}(\Lambda_{ct}+\Lambda_l)u_t$,
where  $\Lambda_{ct}$ stands for the counter term and $\Lambda_l$
arises from the one-loop vertex correction.
Following Refs.~\cite{kadeer, Czarnecki:1992ig, Liu:1992qd},  the counter term
of the vertex includes the mass and the wave-function renormalizations of both the top and bottom quarks  as
\begin{eqnarray}
\Lambda_{ct}=(a+b) \bigg(\frac{\delta Z_b}{2}+\frac{\delta Z_t}{2}-
\frac{\delta m_t}{m_t}\bigg)\frac{1+\gamma_5}{2}
+(a-b) \bigg(\frac{\delta Z_b}{2}+\frac{\delta Z_t}{2}-\frac{\delta m_b}{m_b}\bigg)\frac{1-\gamma_5}{2},
\end{eqnarray}
where, the mass and  the wave function renormalization constants  read
\begin{eqnarray}\label{mass}
\frac{\delta m_q}{m_q}&=&\frac{\alpha_s(\mu_R)}{4\pi}C_F\bigg(\frac{3}{\epsilon_{UV}}-3\gamma_E+
3\ln\frac{4\pi\mu_F^2}{m_q^2}+4\bigg),
\nonumber\\
\delta Z_q &=& -\frac{\alpha_s(\mu_R)}{4\pi}C_F\bigg(\frac{1}{\epsilon_{UV}}+\frac{2}{\epsilon_{IR}}
-3\gamma_E+3\ln\frac{4\pi\mu_F^2}{m_q^2}+4\bigg).
\end{eqnarray}
Here,  $m_q$ is the mass of the relevant quark and  $C_F=(N_c^2-1)/(2N_c)=4/3$ for $N_c=3$ quark colors.
In the equation  above, $\epsilon_{IR}$ and $\epsilon_{UV}$ represent infrared and
ultraviolet singularities and $\gamma_E=0.577216\cdots$ stands for the Euler constant.\\
The real part of the one-loop vertex correction reads
\begin{eqnarray}
\Lambda_l=\frac{\alpha_s m_t^2}{\pi}C_F\big[(a^2-b^2)\sqrt{R}G_++(a^2+b^2)G_-\big],
\end{eqnarray}
with
\begin{eqnarray}
G_+&=&4m_t^2 p_0 C_0(m_b^2,m_t^2,m_{H^+}^2,m_b^2,0,m_t^2)
+B_0(m_b^2,0,m_b^2)+2B_0(m_{H^+}^2,m_b^2,m_t^2)
\nonumber\\
&&+B_0(m_t^2,0,m_t^2)-2,
\nonumber\\
G_-&=&4m_t^2 p_0^2 C_0(m_b^2,m_t^2,m_{H^+}^2,m_b^2,0,m_t^2)
+(2p_0-R)B_0(m_b^2,0,m_b^2)
\nonumber\\
&&+(1+R)B_0(m_{H^+}^2,m_b^2,m_t^2)
+(2p_0-1)B_0(m_t^2,0,m_t^2)-2p_0,
\end{eqnarray}
where, $B_0$ and $C_0$ functions are the Passarino-Veltman 2-point and 3-point integrals which
 can be found in Ref.~\cite{Dittmaier:2003bc}.\\
All uv-singularities  are canceled after summing all virtual corrections up but the ir-divergences
are remaining which are now labeled by $\epsilon$.
Putting everything together, the virtual differential decay rate normalized to the Born total width, reads
\begin{eqnarray}\label{virt}
\frac{1}{\tilde\Gamma_0}\frac{d\tilde\Gamma^{\textbf{vir}}_b}{dx_b}&=&
\frac{\alpha_s(\mu_R)}{2\pi}C_F
\delta(1-x_b)\bigg\{-2+\ln R\bigg[\frac{2(1-p_0)}{y}
-\frac{3ab p_0}{H}-\frac{2 p_0}{p_3}(Y_p+Y_w)\bigg]
\nonumber\\
&&-\frac{2p_0}{p_3}Y_p\bigg(Y_p-\ln y
-\frac{p_3^2}{yp_0H}\big[2\sqrt{R}(a^2-b^2)+(a^2+b^2)(1+R)\big]\bigg)
\nonumber\\
&&-2\bigg[1-\frac{p_0}{p_3}Y_p\bigg]\bigg(\ln\frac{4\pi \mu_F^2}{m_t^2}-
\gamma_E+\frac{1}{\epsilon}\bigg)
-\frac{2p_0}{p_3}\bigg[Li_2(p_-)-Li_2(p_+)+Li_2(1-\frac{p_-}{p_+})\bigg]\bigg\},
\nonumber\\
\end{eqnarray}
where, $Li_2(x)=-\int_0^x(dt/t)\ln(1-t)$ is the Spence function.
In Eq.~(\ref{virt}), one has
\begin{eqnarray}
 ab=\frac{G_F}{\sqrt{2}}|V_{tb}|^2(m_t^2-m_b^2)\cot^2\beta,
 \end{eqnarray}
 in model I, and
 \begin{eqnarray}
ab=\frac{G_F}{\sqrt{2}}|V_{tb}|^2(m_t^2\cot^2\beta-m_b^2\tan^2\beta),
\end{eqnarray}
in model II. The terms $a^2+b^2$ and $a^2-b^2$ are given in Eqs. (\ref{model1}) and
 (\ref{model2}) in both models.

\boldmath
\subsection{Real one-loop corrections (Bremsstrahlung)}\label{real}
\unboldmath

The  ${\cal O}(\alpha_S)$ real gluon emission  amplitude reads
\begin{eqnarray}
M^{tree}=g_s\frac{\lambda^a}{2}\bar u(p_b, s_b)\big\{\frac{2p_t^\mu-
\displaystyle{\not}p_g \gamma^\mu}{2p_t \cdot p_g}
-\frac{2p_b^\mu+\gamma^\mu \displaystyle{\not}p_g}
{2p_b\cdot p_g}\big\}(a+b\gamma_5) u(p_t, s_t)\epsilon_{\mu}^{\star}(p_g,r),
\nonumber\\
\end{eqnarray}
where the polarization vector of the gluon  is denoted by $\epsilon(p_g, r)$.
As before, to regulate the IR-divergences we work in $D=4-2\epsilon$  dimensions and for simplicity,
we choose the top quark rest-frame. To get the correct finite terms in the normalized
differential decay rate, the Born width ${\tilde\Gamma}_0$  will have to be evaluated in
the dimensional regularization at ${\cal O}(\epsilon)$, i.e.
$\tilde\Gamma_0\rightarrow \tilde\Gamma_0\{1-\epsilon
\big[2\ln(2p_3)+\gamma_E-\ln(4\pi\mu^2/m_t^2)-2\big]\}$.
When one integrates over the phase space for the real-gluon radiation,  terms of the
 form $(1-x_b)^{-1-2\epsilon}$ arise which are due to  the radiation of a soft gluon in top decay.
 Therefore, for a massive b quark, where $x_{b,min}=\beta$, we use the following expansion
 \begin{eqnarray}
\frac{(x_b-\beta)^{2\epsilon}}{(1-x_b)^{1+2\epsilon}}=-\frac{1}{2\epsilon}\delta(1-x_b)+
\frac{1}{(1-x_b)_+}+{\cal O}(\epsilon),
 \end{eqnarray}
where the plus distribution is  defined as
\begin{eqnarray}\label{plus}
\int_\beta^1 \frac{ f(x_b)}{(1-x_b)_+} dx_b=\int_\beta^1 \frac{f(x_b)-f(1)}{1-x_b} dx_b
+f(1)\ln(1-\beta).
\end{eqnarray}

\boldmath
\subsection{Parton-level results for $d\tilde\Gamma/dx_a$ in FFN scheme}
\unboldmath

Now, we present our analytic results for partial decay rate normalized to the Born width
 in the FFN scheme, by summing the tree level, the virtual and the real contributions. Our  result reads
\begin{eqnarray}\label{first}
\frac{1}{\tilde\Gamma_0}\frac{d\tilde\Gamma_b}{dx_b}&=&\delta(1-x_b)+
\frac{C_F\alpha_s(\mu_R)}{\pi}\Bigg\{
\delta(1-x_b)\bigg[\frac{(1-R)Y_w}{p_3}-2\ln\frac{2p_0 p_+}{\sqrt{y}}-1
\nonumber\\
&&-2\frac{p_0}{p_3}\big[Li_2(p_-)-Li_2(p_+)+Li_2(1-\frac{p_-}{p_+})\big]
+2\frac{Y_p}{p_3}\bigg(p_3+\frac{p_0-R}{2}+p_0\ln\frac{2p_0\sqrt{y}}{p_+}+
\nonumber\\
&&\frac{p_3^2}{2yH}\big[2(a^2-b^2)\sqrt{R}+(1+R)(a^2+b^2)\big]\bigg)
+\big(\frac{1+y-p_0}{y}-\frac{p_0}{p_3}Y_w-\frac{3p_0}{2H}ab\big)\ln R\bigg]
\nonumber\\
&&-\frac{2(T+x_b \Phi(x_b))}{H p_3(1-x_b)_+}\big[\sqrt{R}(a^2-b^2)+(a^2+b^2)p_0x_b\big]
-p_0\frac{a^2+b^2}{H p_3}\big[T+(1+x_b)\Phi(x_b)\big]\Bigg\}.\nonumber\\
\end{eqnarray}
Integrating $d\tilde\Gamma_b/dx_b$ of Eq. (\ref{first}) over $x_b(\beta<x_b<1)$,
we obtain the total decay rate presented in Refs.~\cite{kadeer, Czarnecki:1992ig}.

Since, the B-meson can be also produced from the fragmentation of the emitted real gluon,
we also need the differential decay rate $d\tilde\Gamma_g/dx_g$ in
the FFN scheme, where $x_g=E_g/(m_t p_0)$ is the scaled energy fraction of the real gluon.
To calculate the $d\tilde\Gamma_g/dx_g$, we integrate over the momentum of b-quark by
 fixing the momentum of gluon in the phase space. Our result is listed here
\begin{eqnarray}\label{second}
\frac{1}{\tilde\Gamma_0}\frac{d\tilde\Gamma_g}{dx_g}&=&
\frac{C_F\alpha_s(\mu_R) p_0^2(1-x_g)}{\pi p_3 x_g H}\bigg\{
\bigg[(a^2+b^2)\big(1-x_g+\frac{1}{1-x_g}\big)+2\beta(a^2-b^2)\bigg]
\nonumber\\
&&\times\tanh^{-1}\sqrt{1-\frac{\beta^2(1-2p_0x_g)}{(1-x_g)^2}}
-\bigg[(a^2+b^2)\bigg(2+\frac{p_0x_g^2\big[2-p_0(1+3x_g)\big]}{(1-2p_0 x_g)^2}\bigg)
\nonumber\\
&&+2\beta(a^2-b^2)\bigg] \sqrt{1-\frac{\beta^2(1-2p_0x_g)}{(1-x_g)^2}}\bigg\}.
\end{eqnarray}

\boldmath
\section{GM-VFN scheme}
\label{sec:two}
\unboldmath

Our main purpose is to calculate the scaled-energy distribution of the B-hadron produced in
the inclusive process $t\rightarrow BH^++X$ in the 2HDM, where  $X$ stands for the unobserved final state.
Thus we calculate the partial decay width  of the corresponding process
differential in $x_B$ ( $d\Gamma/dx_B$), at NLO in the GM-VFN scheme, where $x_B=E_B/(m_t p_0)$ is
the scaled energy fraction of the B-hadron.
In the top quark rest frame, the B-hadron has energy $E_B=p_t\cdot p_B/m_t $, where
$m_B\le E_B\le [m_t^2+m_B^2-m^2_{H^+}]/(2m_t) $. In the case of gluon fragmentation, $g\rightarrow B$,
it has energy $m_B\le E_B\le [m_t^2+m_B^2-(m_b+m_{H^+})^2]/(2m_t) $.\\
According to the factorization  theorem of QCD \cite{collins}, the B-hadron energy
distribution  can be obtained by  the convolution of the parton-level spectrum with
the nonperturbative fragmentation function $D_a^B(z, \mu_F)$,
\begin{equation}\label{eq:master}
\frac{d\Gamma}{dx_B}=\sum_{a=b, g}\int_{x_a^\text{min}}^{x_a^\text{max}}
\frac{dx_a}{x_a}\,\frac{d\Gamma^{GM}_a}{dx_a}(\mu_R, \mu_F) D_a^B(\frac{x_B}{x_a}, \mu_F),
\nonumber\\
\end{equation}
where, $\mu_F$ is the factorization scale and $\mu_R$ is the renormalization scale which  is related to
 the renormalization of the QCD coupling constant.
A choice often made is to set  $\mu_R=\mu_F$ and we shall use this
convention for most of the results.
Here,  $d\Gamma^{GM}_a/dx_a$ is the differential decay width of the parton-level process $t\to a+X$ at
NLO in the GM-VFN scheme, where X comprising the $H^+$ boson and any other parton.
 We now discuss the evaluation of the quantities
$d\Gamma^{GM}_a(\mu_R, \mu_F) /dx_a$  in the GM-VFN scheme in detail.\\
In Ref.~\cite{MoosaviNejad:2011yp},  using  the ZM-VFN scheme we  evaluated  the quantities
 $1/\Gamma_0\times d\hat\Gamma_a/dx_a (a=b,g)$ for the
 process $t\to a+X$, where $m_b=0$  is put right from the beginning.
In this scheme, $m_b$ only sets the initial
scale $\mu_F^{ini}={\cal O}(m_b)$ of the DGLAP evolution, however all information
on the $m_b$ dependence of $d\hat\Gamma_a/dx_a$  is wasted.\\
In sec.~\ref{sec:one} of the present paper, we applied the FFN scheme which
contains of the full $m_b$ dependence.  In this scheme,
the large logarithmic singularities of the type $(\alpha_s/\pi)\ln R$, where $R=m_b^2/m_t^2$,  spoil the
convergence of the perturbative expansion when $m_b/m_t\rightarrow 0$ (see Eq.~(\ref{first})). The GM-VFN
scheme is devised to resum the large logarithms in $m_b$ and to retain the whole
nonlogarithmic $m_b$ dependence at the same time. This is achieved by introducing
appropriate subtraction terms in the NLO FFN expressions for $d\tilde\Gamma_a/dx_a$,
so that the NLO ZM-VFN results are exactly recovered in the limit $m_b/m_t\rightarrow 0$.
With this explanation, the subtraction terms are constructed as
\begin{eqnarray}
\frac{1}{\Gamma_0}\frac{d\Gamma^{Sub}_a}{dx_a}=\lim_{m_b\rightarrow 0}
\frac{1}{\Gamma_0}\frac{d\tilde\Gamma_a^{FFN}}{dx_a}-
\frac{1}{\Gamma_0}\frac{d\hat\Gamma_a^{ZM}}{dx_a},
\end{eqnarray}
and the GM-VFN results are obtained by subtracting the subtraction terms from the FFN ones \cite{Kniehl:2, Kniehl:3}, as
\begin{eqnarray}
\frac{1}{\Gamma_0}\frac{d\Gamma^{GM}_a}{dx_a}=
\frac{1}{\Gamma_0}\frac{d\tilde\Gamma_a^{FFN}}{dx_a}-
\frac{1}{\Gamma_0}\frac{d\Gamma^{Sub}_a}{dx_a}.
\end{eqnarray}
Taking the limit $m_b\rightarrow 0$ in Eqs. (\ref{first}) and (\ref{second}), we obtain  the subtraction terms as
\begin{eqnarray}\label{bff}
\frac{1}{\Gamma_0}\frac{d\Gamma^{Sub}_b}{dx_b}&=&\frac{\alpha_s(\mu_R)}{2\pi}C_F
\bigg\{\frac{1+x_b^2}{1-x_b}\bigg[\ln\frac{\mu_F^2}{m_b^2}-2\ln(1-x_b)-1\bigg]\bigg\}_+,
\nonumber\\\\
\frac{1}{\Gamma_0}\frac{d\Gamma^{Sub}_g}{dx_g}&=&\frac{\alpha_s(\mu_R)}{2\pi}C_F
\frac{1+(1-x_g)^2}{x_g}\bigg(\ln\frac{\mu_F^2}{m_b^2}-2\ln x_g-1\bigg).
\nonumber\\
\end{eqnarray}
As it is guaranteed by Collin's factorization theorem \cite{collins},  the subtraction terms are universal
 and as we already presented in Ref. \cite{Kniehl:2012mn}, Eq. (\ref{bff}) coincides with the perturbative FF of the
 transition $b\rightarrow b$ \cite{Mele:1990cw}.

\boldmath
\section{Numerical results}
\label{sec:three}
\unboldmath

Now we present  our phenomenological predictions  by performing a numerical analysis.
As it is referred in Ref.~\cite{Ali:2009sm},
a charged Higgs having a mass in the range $80 GeV\leq m_{H^\pm}\leq 160 GeV$ is
a logical possibility
and its effects should be searched for in the decay modes $t\rightarrow  bH^+\rightarrow B\tau^+\nu_\tau+X$.
A beginning along these lines has already been made at the Tevatron \cite{Abbott:1999eca,Abulencia:2005jd},
 but a definitive search  of the charged-Higsses over a good part
of the ($m_{H^+}-\tan\beta$) plane is a program that still has to be carried out and this belongs to the LHC
experiments \cite{Aad:2008zzm}.\\
Following Ref.~\cite{Nakamura:2010zzi}, we adopt
the present lower limit $m_{H^+}>79.3$ GeV obtained from LEP. \\
From Ref.~\cite{Nakamura:2010zzi}, we use the input parameter values
$G_F = 1.16637\times10^{-5}$~GeV$^{-2}$,
$m_t = 172.0$~GeV,
$m_b = 4.90$~GeV,
$m_B = 5.279$~GeV, and
$|V_{tb}|=0.999152$.
We evaluate $\alpha_s^{(n_f)}(\mu_R)$ at NLO in the $\overline{\text{MS}}$
scheme using
\begin{eqnarray}\label{alpha}
\alpha^{(n_f)}_s(\mu)=\frac{1}{b_0\log(\mu^2/\Lambda^2)}
\Big\{1-\frac{b_1 \log\big[\log(\mu^2/\Lambda^2)\big]}{b_0^2\log(\mu^2/\Lambda^2)}\Big\},
\nonumber\\
\end{eqnarray}
with $b_0$ and $b_1$ given by
\begin{eqnarray}
b_0=\frac{33-2n_f}{12\pi}, \quad  b_1=\frac{153-19n_f}{24\pi^2},
\end{eqnarray}
where $\Lambda$ is the typical QCD scale and we adopt
$\Lambda_{\overline{\text{MS}}}^{(5)}=231.0$~MeV adjusted such
 that $\alpha_s^{(5)}=0.1184$ for $m_Z=91.1876$~GeV \cite{Nakamura:2010zzi}.
 In Eq. (\ref{alpha}), $n_f$ is the number of active quark flavors.\\
To describe the transitions $b, g\rightarrow B$, we employ the
realistic nonperturbative $B$-hadron  FFs determined at NLO in
the zero-mass scheme through a global fit  to
$e^+e^-$-annihilation data presented by ALEPH \cite{Heister:2001jg} and OPAL
\cite{Abbiendi:2002vt} at CERN LEP1 and by SLD \cite{Abe:1999ki} at SLAC SLC.
Specifically,  for the $b\to B$ transition the power model  $D_b(z,\mu_F^\text{ini})=Nz^\alpha(1-z)^\beta$
was used at the initial scale $\mu_F^\text{ini}=4.5$~GeV, while the  light-quark and gluon FFs were generated
via the DGLAP evolution.
The result of fit read $N=4684.1$, $\alpha=16.87$, and $\beta=2.628$ \cite{Kniehl:2008zza}.
Note, if the same experimental data are fitted in the ZM-VFN and GM-VFN schemes,
 the resulting FFs will be somewhat different. As it is shown in Ref.~\cite{do}, the hadronization
 of the bottom quark is identified to be
 the largest source of uncertainty in the measurement of the top quark mass.

\begin{figure}
\begin{center}
\includegraphics[width=0.7\linewidth,bb=20 192 552 629]{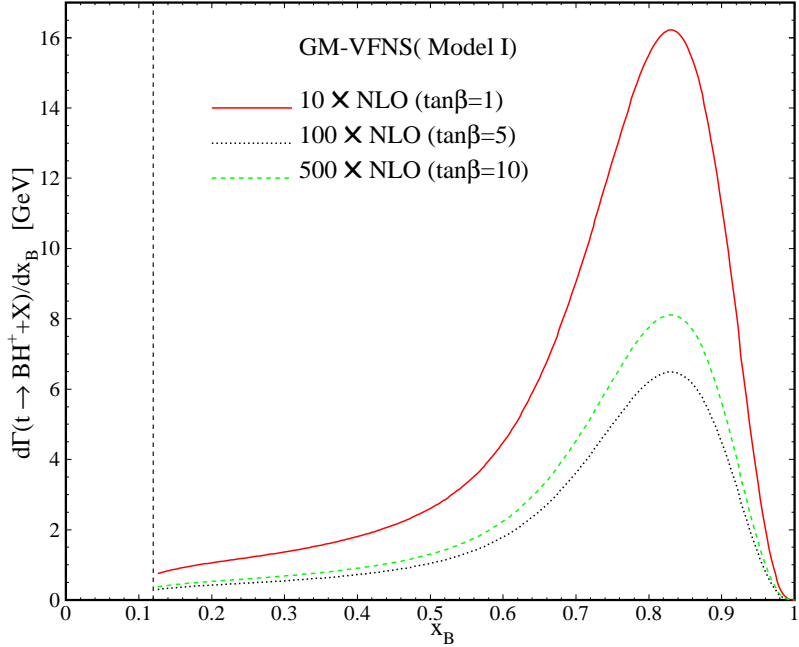}
\caption{\label{fig1}%
$d\Gamma(t\to BH^++X)/dx_B$ as a function of $x_B$ in the GM-VFN
scheme for model I, taking $m_{H^+}=120$  GeV and  $\tan\beta=1,5$ and $10$.}
\end{center}
\end{figure}

\begin{figure}
\begin{center}
\includegraphics[width=0.7\linewidth,bb=20 192 552 629]{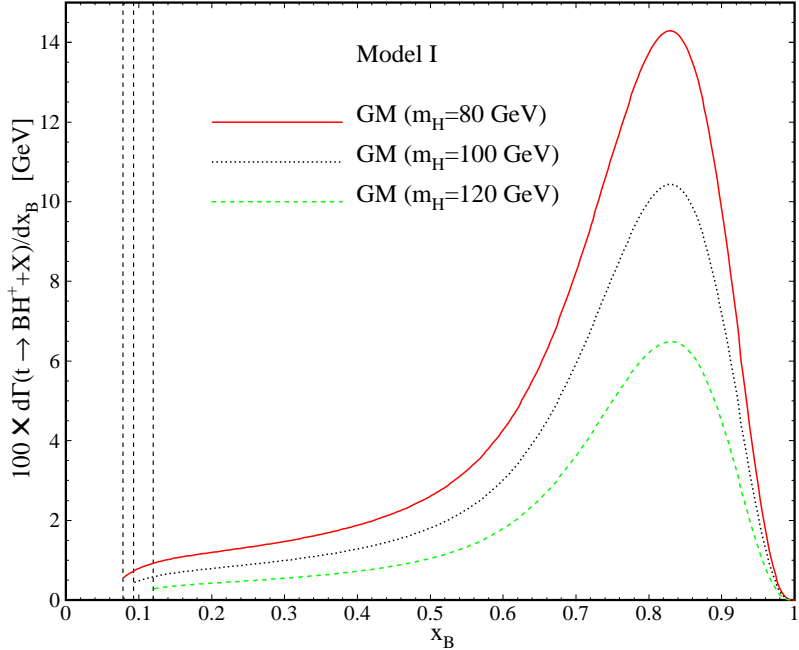}
\caption{\label{fig2}%
$d\Gamma(t\to BH^++X)/dx_B$ as a function of $x_B$ for model I. Different values of the Higgs
boson mass is considered, i.e. $m_{H^+}=80, 100$ and 120 GeV. Other free
 parameter is fixed to $\tan\beta=5$.}
\end{center}
\end{figure}

To present our results for the scaled-energy ($x_B$ ) distribution of B-hadrons, we
consider the quantity $d\Gamma(t\to BH^++X)/dx_B$
taking the $H^+$ boson to be stable. In Fig.~\ref{fig1}, we show our prediction for the
size of the NLO corrections in the GM-VFN scheme for the model I. Here, the mass of
Higgs boson is fixed to $m_{H^+}=120$  GeV and the different values of $\tan\beta$
are considered, i.e. $\tan\beta=1,5$ and $10$.
However, as in Ref.~\cite{Schael:2006cr} it is pointed out, the small values of $\tan\beta$ are  excluded by
the indirect limits in the ($m_{H^\pm}, \tan\beta$) plane.  For example, taking the
CP-conserving scenario $m_h$-max and a top quark mass of
$174.3$ $ GeV$,  values of $\tan\beta$ between 0.7 and 2.0
are excluded, but this range depends considerably on the
assumed top quark mass and may also depend on $M_{SUSY}$ (the soft  SUSY breaking
scale parameter).\\
 In Fig.~\ref{fig1}, Both the b-quark and gluon fragmentations are included.
 In Ref.~\cite{MoosaviNejad:2011yp}, we showed
the $g\rightarrow B$ contribution is negative and appreciable only in the low-$x_B$ region. For higher
values of $x_B$ the NLO result is practically exhausted by the $b\rightarrow B$ contribution.\\
From Fig.~\ref{fig1}, it can be seen that when $\tan\beta$ is
increased the decay rate is decreased and the peak position is shifted towards higher values of
$x_B$.  In Ref.~\cite{kadeer},  it is shown when
the values of $\tan\beta$ exceed $\tan\beta=2$,
the decay rate becomes quite small. Here, the mass of B-hadron creates a
threshold at $x_B=2m_B/(m_t(1+R-y))=0.12$.

\begin{figure}
\begin{center}
\includegraphics[width=0.7\linewidth,bb=20 192 552 629]{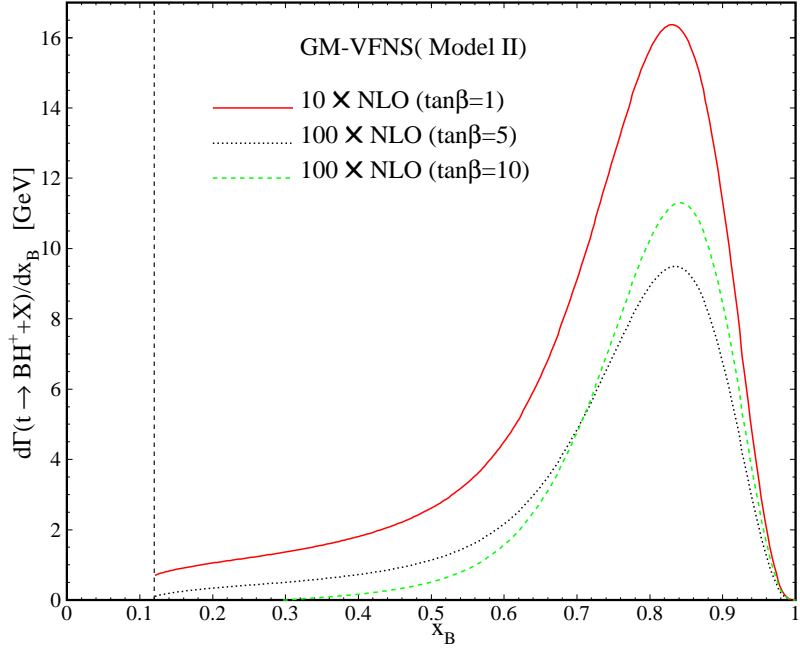}
\caption{\label{fig3}%
$x_B$ spectrum in top decay considering the decay mode $t\rightarrow BH^++X$,
taking  $m_{H^+}=120$  GeV
 and  $\tan\beta=1,5$ and $10$ in model II. Threshold at $x_B$ is shown.}
\end{center}
\end{figure}

\begin{figure}
\begin{center}
\includegraphics[width=0.7\linewidth,bb=20 192 552 629]{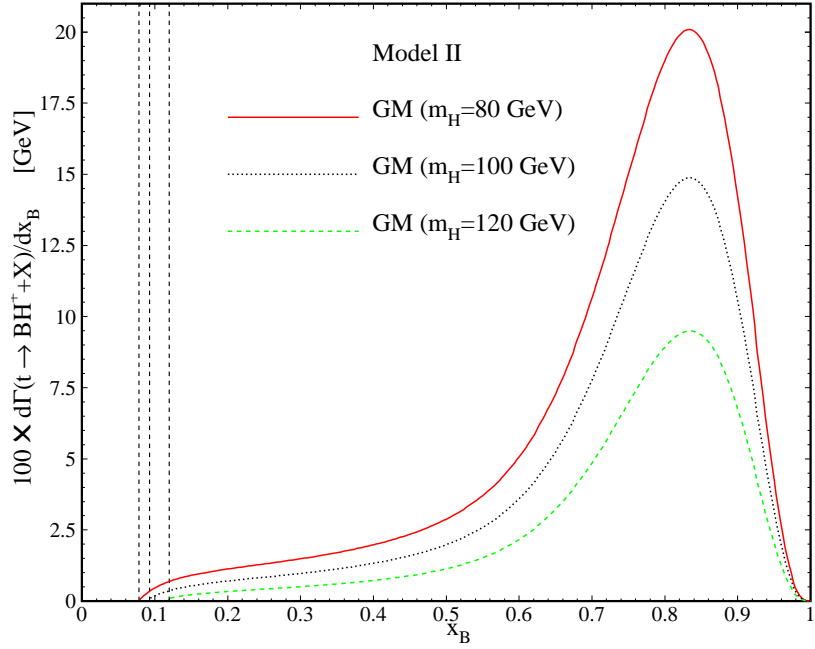}
\caption{\label{fig4}%
$x_B$ spectrum for model II considering the different values of the Higgs boson mass,
 i.e. $m_{H^+}=80, 100$ and 120 GeV, by fixing  $\tan\beta=5$.}
\end{center}
\end{figure}

\begin{figure}
\begin{center}
\includegraphics[width=0.7\linewidth,bb=20 192 552 629]{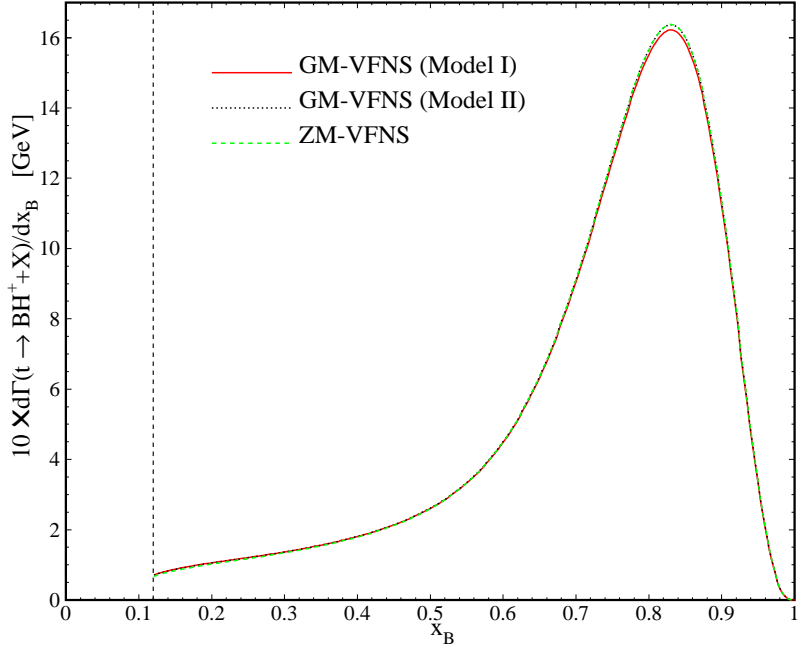}
\caption{\label{fig5}%
$d\Gamma(t\rightarrow B+H^+)/dx_B$ as a function of  $x_B$ at NLO. The GM-VFNs
results in two models are compared to the  ZM-VFN scheme using $m_{H^+}=120$  GeV and $\tan\beta=1$.}
\end{center}
\end{figure}

\begin{figure}
\begin{center}
\includegraphics[width=0.7\linewidth,bb=20 192 552 629]{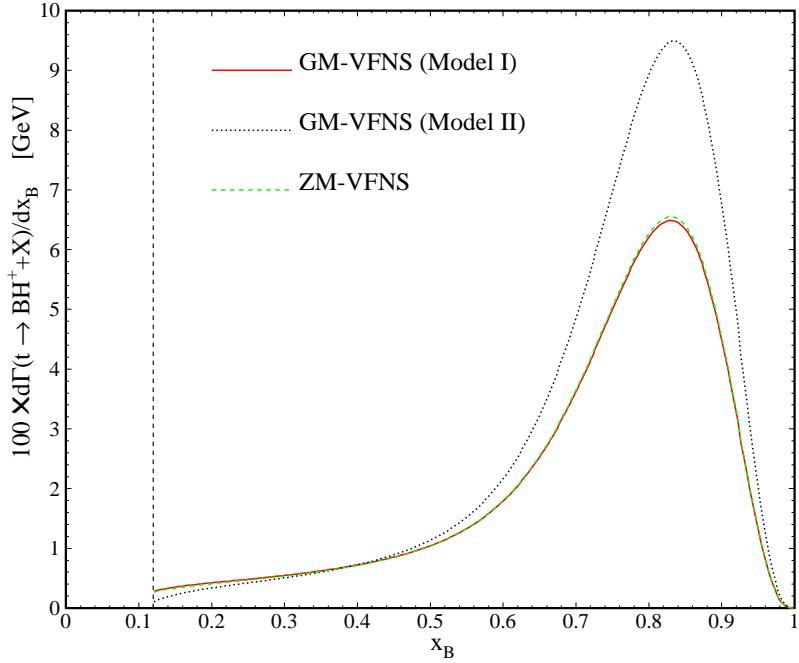}
\caption{\label{fig6}%
As in Fig.~\ref{fig5}, but using  $\tan\beta=5$.}
\end{center}
\end{figure}

\begin{figure}
\begin{center}
\includegraphics[width=0.8\linewidth,bb=20 192 552 629]{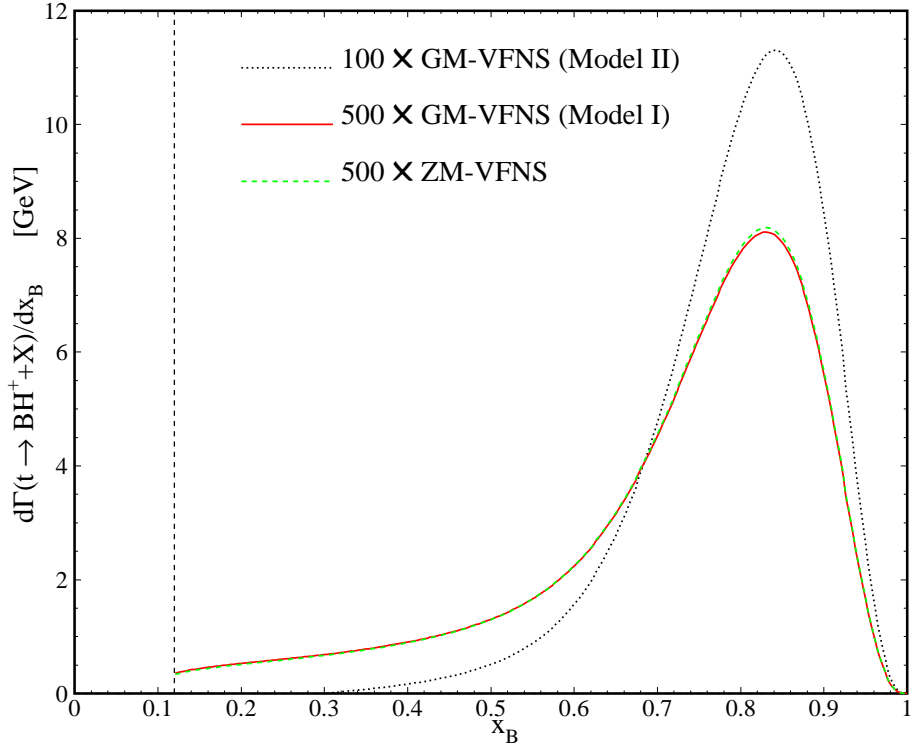}
\caption{\label{fig7}%
$x_B$ spectrum at NLO as Figs.~\ref{fig5} and \ref{fig6} but using $\tan\beta=10$.}
\end{center}
\end{figure}

\begin{figure}
\begin{center}
\includegraphics[width=0.8\linewidth,bb=20 192 552 629]{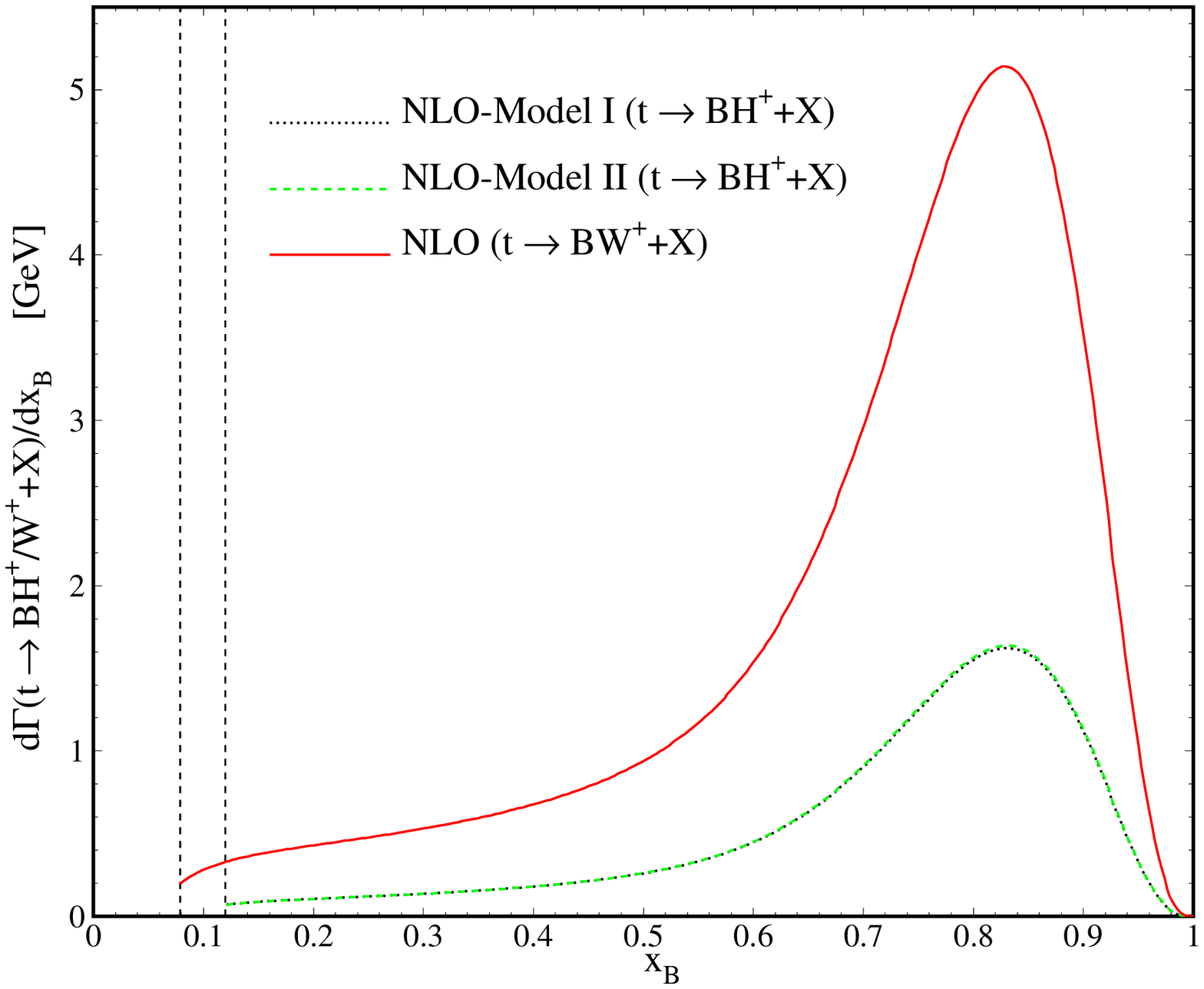}
\caption{\label{fig8}%
$x_B$ spectrum in top decay considering the decay modes  $t\rightarrow BW^++X$ (solid line) and
$t\rightarrow BH^++X$  (dashed and dotted lines), taking $m_{W^+}=80.399$ GeV, $m_{H^+}=120$  GeV and $\tan\beta=1$.}
\end{center}
\end{figure}

Adopting the limit $80 GeV\leq m_{H^\pm}\leq 160 GeV$ from Ref.~\cite{Ali:2009sm},
in Fig.~\ref{fig2}  we study the energy spectrum of the B-hadron for model I  in
different values of the Higgs boson mass, i.e. $m_{H^+}=80, 100$ and 120 GeV,
by fixing  $\tan\beta=5$. As mentioned,  the mass of B-hadron is responsible for the thresholds
at $x_B=0.08$ (for $m_{H^+}=80$ GeV), $x_B=0.09$ (for $m_{H^+}=100$ GeV)
and $x_B=0.12$ (for $m_{H^+}=120$ GeV).

In Fig.~\ref{fig3} and Fig.~\ref{fig4},  the results presented in Figs.~\ref{fig1}
and \ref{fig2} are regained for model II. The thresholds are as before.

In Fig.~\ref{fig5}, we compared the energy spectrum of B-hadron in the
GM-VFN ($m_b\neq 0$) and ZM-VFN ($m_b=0$) schemes, using $\tan\beta=1$ and
$m_{H^+}=120$ GeV. As it is shown, the results
of both models are the same in the GM-VFN scheme but the result of ZM-VFN
scheme shows an enhancement in the size of decay rate  about $1.3\%$ at $x_B=0.8$.
In Figs.~\ref{fig6} and \ref{fig7}, this comparison is done  using $\tan\beta=5, 10$ and $m_{H^+}=120$ GeV.
As it is seen,  the prediction for the
energy spectrum extremely depends on the model at the large values of  $\tan\beta$ when $m_b\neq 0$.
Therefore, our most reliable prediction for the
energy spectrum  is made at NLO in the GM-VFN  scheme.\\
In Fig.~\ref{fig8}, the energy spectrum of B-hadron in decay modes $t\rightarrow BW^++X$ and $t\rightarrow BH^++X$
are compared. As before, the mass of B-hadron create  the thresholds
at $x_B=0.12$ (for $m_{H^+}=120$ GeV) and $x_B=0.08$ (for $m_{W^+}=80.399$ GeV).
It is obvious that  the contribution of the top decay
mode in the SM is always larger than the one coming from  the 2HDM.
However, to obtain the total energy spectrum of B-hadron in the top
quark decay all decay modes including $t\rightarrow B+W^+/H^+$ should be summed up.

\boldmath
\section{Conclusions}
\label{sec:four}
\unboldmath

Clearly, the decay modes $t\rightarrow W^++B$  have been and will be the prime source of
information on the top quark mass. We have studied these dominant decay modes 
 along the lines of Ref.~\cite{Kniehl:2012mn}.
In the theories beyond the standard model including the two-Higgs-doublet,
the top quarks also decay into a charged Higgs and a bottom quark thus
it may be useful to also use $t\rightarrow H^++B$  events for a cross check.
The search for the light charged Higgs boson ($m_{H^+}<m_t$) produced from the decay mode
 $t\rightarrow bH^+(\rightarrow \tau^+\nu)$
has been performed using about $1 fb^{-1}$ of data collected in proton-proton collisions at
$\sqrt{s}=7$ TeV \cite{Djouadi:2005gj}.\\
 In our previous work \cite{MoosaviNejad:2011yp},  we applied the ZM-VFN scheme to study
 the dominant decay channel
 $t\rightarrow BH^++X$ in the 2HDM, where the mass of bottom-quark was set to zero and thus all
 information on the $m_b$ dependence of the B-hadron spectrum
was wasted. In the present work, we studied the energy spectrum of B-hadron in top decay
considering  the quantity $d\Gamma/dx_B$ in GM-VFN scheme. Our main purpose was to
investigate both the effect of b-quark mass and the gluon fragmentation to the B-hadron energy
distribution. In order to study these effects  we have calculated, for the first time, an analytic
expression for  the NLO radiative corrections to the differential  top decay width
$d\tilde\Gamma/dx_a (a=b, g)$ in two variants of the 2HDM.  To ensure our results, we  have checked
that by integrating $d\tilde\Gamma/dx_b$ over $x_b$ we recover
 the known results presented in Refs.~ \cite{kadeer, Czarnecki:1992ig}.\\
In Ref.~\cite{MoosaviNejad:2011yp}, we showed in the limit $m_b\rightarrow 0$ our results
 in both models are the same
but in the present work we have checked  the results for the energy spectrum of B-hadron
are completely different for two models in the general 2HDM when $m_b\neq 0$ and $\tan\beta$
is large.  In conclusion, the most reliable predictions for  $d\Gamma/dx_B$
is made at NLO in the GM-VFN  scheme.\\
Comparing future measurements of $d\Gamma/dx_B$ at the LHC with the presented predictions will be
important for our understanding of the Higgs coupling in 2HDM and new physics beyond the
standard model. Our results in Refs.~\cite{Kniehl:2012mn, MoosaviNejad:2011yp} and the present work both will be
able to test the universality of  the B-hadron fragmentation functions and
provide a clean method to gauge the normalization of Monte Carlo event generators.

\begin{acknowledgments}
I would like to thank Professor Bernd Kniehl for reading  the manuscript and also for his important comments.
I  would also like to thank Dr Mathias Butenschon  for reading and improving the manuscript.
This work was supported by Yazd university and the Institute for Research in Fundamental
Science (IPM).
\end{acknowledgments}

\end{document}